\documentclass[preprint,12pt]{elsarticle}

\usepackage{amssymb}
\usepackage{amsmath}
\usepackage{enumerate}
\usepackage{lscape}
\usepackage{rotating}
\pdfoutput=1 

\usepackage{lineno}

\journal{Nuclear Instruments and Methods A}

\begin{document}

\begin{frontmatter}



\title{An analytical model for photomultiplier tube calibration}


\author{Leonidas N. Kalousis$^a$}

\affiliation{organization={Physics Department, National Technical University},
            city={Zografou},
            postcode={15780}, 
            state={Athens},
            country={Greece}}

\begin{abstract}

The purpose of the present article is to offer an analytical model that can be used for the calibration of photomultiplier tubes (PMTs). 
The derivation of the mathematical formulae is discussed extensively and we apply this machinery to a real problem; the gain determination of the R7081 Hamamatsu PMT. 
We hope that our work can provide a third, powerful alternative to the task of gain determination.  
We also believe that the techniques explained in this article can be applied to other problems in PMT calibration. 
 
\end{abstract}




\end{frontmatter}


\section{Introduction}
\label{sec:intro}

The standard machinery for photomultiplier tube (PMT) calibration has been presented in an influential paper written by E. H. Bellamy and collaborators~\cite{bellamy}. 
Years later, R. Dossi \emph{et al.} proposed a specific model for the single photoelectron (SPE) response function $S(x)$ that can treat a vast number of PMTs~\cite{dossi}. 
It was based on a carefully chosen combination of an exponential and a truncated gaussian distribution:
\begin{align}
S(x) =    \left( \ w \alpha e^{-\alpha x } + \frac{(1-w)}{g_N} \frac{1}{\sqrt{2\pi}\sigma} e^{ - \frac{( x - Q )^2}{2\sigma^2}} \ \right) H(x).       \label{eq:S}
\end{align}
Note that photoelectrons that backscatter and miss the first amplification stage are modeled in $S(x)$ through the exponential term. 
The slope of the exponential is $\alpha$ and the $w$ pre-factor parametrizes the probability for this process to happen. 
A better discussion of this, with further experimental tests, can be found in ref.~\cite{dossi}. 
$H(x)$ is the Heaviside step function and $Q$ and $\sigma$ are the parameters of the truncated gaussian that models the amplification of the full dynode chain.  
The $g_N$ factor ensures that the truncated gaussian is properly normalized and equals to:
\begin{align}
g_N = \frac{1}{2} \textrm{erfc} \left( - \frac{Q}{\sqrt{2}\sigma} \right), 
\end{align}
where erfc($z$) is the complementary error function~\cite{error}. 

The mathematical formulae involved in the Dossi model are rather complicated and an analytical solution has eluded experimental physicists. 
Up to now, two main approaches are available for the deduction of the charge response function $S_R(x)$.
\begin{enumerate}[i.]
\item  A method that computes the convolution integrals numerically, using a recursive algorithm, 
for the the first photoelectron (PE) peaks and approximates higher order corrections with symmetric gaussians~\cite{dossi}. 
\item A novel technique, based on the Discrete Fourier Transform (DFT), that calculates the charge response function $S_R(x)$ numerically for all orders in the Poisson mean $\mu$~\cite{me}. 
\end{enumerate}
Both procedures give consistent results when carefully applied. 
In this article, we have tried to solve the Dossi model analytically, offering thus a third solution to this task. 
Nonetheless, it is always preferable to have an analytical formula at hand since: 
(a) an exact formula is expected to run faster and 
(b) it is devoid of possible numerical issues that can destabilize the aforementioned methods.  

The paper is organized as follows. 
In Section~\ref{sec:mod} we present in great detail the calculations that lead to the analytical model we put forward. 
It should be emphasized from the onset that, while the mathematics might seem unnecessary and elaborate at first glance, the final equations are rather simple and easy to implement. 
In Section~\ref{sec:sim} we analyze generated data to assess the validity of the model. 
Finally, in Section~\ref{sec:res} we apply this formula to a real problem; the calibration of the Hamamatsu R7081 PMT. 
Attention was paid to showcase that the analytical model provides consistent figures with the already established methods. 
We close this publication with some general remarks concerning further uses of this mathematical scheme.

\section{Analytical model}
\label{sec:mod}

\subsection{General form of   $S^{(n)}_R(x)$ }

The behavior of a PMT, when illuminated with light pulses of constant number of photons, is dictated by the charge response function $S_R(x)$:
\begin{align}
S_R(x) = \sum_{n=0}^{+\infty} P(n;\mu) S^{(n)}_R(x), \label{eq:SR}
\end{align}
where 
\begin{align}
S^{(n)}_R(x) = (S_n*B)(x). 
\end{align}
$ P(n;\mu)$ is the Poisson distribution with mean value $\mu$. 
$S_n(x)$ corresponds to the $n$-times convolution of the SPE response function $S(x)$ and $B(x)$ refers to the pedestal. 
An extensive review describing the main steps in PMT calibration can be found elsewhere~\cite{me}. 
Let us assume, following the paper of Dossi \emph{et~al.}, that the SPE charge amplification can be written as:
\begin{align}
S(x) = w f(x) + (1-w)g(x), \label{eq:S}
\end{align}
where 
\begin{align}
f(x) = \alpha e^{-\alpha x } \ H(x),
\end{align}
and
\begin{align}
g(x) = \frac{1}{g_N} \frac{1}{\sqrt{2\pi}\sigma} \ e^{ - \frac{( x - Q )^2}{2\sigma^2}} \ H(x). 
\end{align}
The principal $S^{(n)}_R(x)$ functions are equal to: 
\begin{align}
S^{(n)}_R(x) & = \sum_{m=0}^{n}  \frac{n!}{m!(n-m)!} \ w^m (1-w)^{n-m} (f_m*g_{n-m} *B )(x)\nonumber \\
                     & = (1-w)^n (g_{n} *B )(x) \nonumber \\
                    & + \sum_{m=1}^{n}  \frac{n!}{m!(n-m)!} \ w^m (1-w)^{n-m} (f_m*g_{n-m} *B )(x), \label{eq:snr1}
\end{align}
where the $f_m(x)$ and $g_{n-m}(x)$ are the $m$-times and $(n-m)$-times convolutions of $f(x)$ and $g(x)$ respectively. 
In this last step we made use of the binomial expansion for $S_n(x)$ \cite{error}. 
The formula can be simplified by setting:
\begin{align}
G_n(x) = (g_{n} *B )(x), 
\end{align}
and 
\begin{align}
h_{m,n}(x) = (f_{m} * G_{n-m} )(x).  
\end{align}
Like this eq.~\eqref{eq:snr1} can be written as:
\begin{align}
S^{(n)}_R(x) & = (1-w)^n G_n(x) + \sum_{m=1}^{n}  \frac{n!}{m!(n-m)!} \ w^m (1-w)^{n-m} h_{m,n}(x). \label{eq:snr2}
\end{align}
We proceed to calculate each term in eq.~\eqref{eq:snr2} separately.

\subsection{Calculation of $S^{(0)}_R(x)$ and $S^{(1)}_R(x)$}

The zeroth term of $S^{(n)}_R(x)$ can be computed easily. 
The result is:
\begin{align}
S^{(0)}_R(x)  = \ & B(x) \nonumber \\
                      = \ & \frac{1}{\sqrt{2\pi}\sigma_0} \ e^{-\frac{(x-Q_0)^2}{2\sigma_0^2}}.
\end{align}
We remind the reader that $f_0(x)=g_0(x)=\delta (x)$. 
This is the well-known pedestal distribution and $Q_0$ and $\sigma_0$ are the mean value and standard deviation of $B(x)$ respectively. 
On the other hand, the $n=1$ term receives two contributions:
\begin{align}
S^{(1)}_R(x)  = w (f*B)(x) + (1-w)(g*B)(x). 
\end{align}                     
We begin with the formula of $(f*B)(x)$ and then write down $(g*B)(x)$.

The convolution between $f(x)$ and $B(x)$ is given by the equation:
\begin{align}
(f*B)(x)  = \ \frac{\alpha}{2} e^{\frac{\alpha^2\sigma_0^2}{2}} e^{-\alpha (x-Q_0)} \ \text{erfc}\left( \frac{Q_0 + \alpha\sigma_0^2 - x }{\sqrt{2}\sigma_0} \right).
\end{align}  
On the other hand, $(g*B)(x)$ equals to:
\begin{align}
(g*B)(x)  = \ \frac{1}{2}  \frac{1}{g_N}  \frac{1}{\sqrt{2\pi}} \frac{1}{\sqrt{\sigma_0^2 + \sigma^2}}   \ e^{ -\frac{ (x-Q_0-Q)^2 }{2(\sigma_0^2 + \sigma^2)}} 
\ \text{erfc}\left(  \frac{ Q_0\sigma^2 -Q\sigma_0^2 -x \sigma^2  }{\sqrt{2} \sigma_0\sigma\sqrt{\sigma_0^2 + \sigma^2} }   \right).
\end{align}  
The proof of both these formulae is challenging and interesting in its own right and it is presented in the \ref{app:sr1} with great care. 

\subsection{Gaussian approximation of $G_n(x)$}

A direct calculation of $G_n(x)$, 
\begin{align}
G_n(x)  = (g_n*B)(x),
\end{align}   
for $n\geq 2$ appears to be both difficult and complicated. 
Perhaps the mathematics can be worked out in some systematic way, but the resulting equations ought to be elaborate to be implemented efficiently in an analysis software. 
For the reasons above, we took the decision to approximate all truncated gaussians in $G_n(x)$ with perfectly symmetric ones. 
In this way $G_n(x)$ becomes:
\begin{align}
G_n(x)  = \ \frac{1}{\sqrt{2\pi} \sigma_n } \ e^{ -\frac{ (x-Q_n)^2 }{2\sigma_n^2} },
\end{align}   
where
\begin{align}
Q_n  & = Q_0 + n Q_g, \\
\sigma_n^2 & = \sigma_0^2 + n \sigma_g^2.  
\end{align}   
Since several convolutions are involved, the truncation of $g(x)$ at negative values of charge should have little impact on the final result.  
$Q_g$ and $\sigma_g$ are the mean value and standard deviation of $g(x)$ and their exact formulae can be found in ref.~\cite{me2}.

\subsection{$h_{m,n}(x)$ in terms of hypergeometric functions}

The major difficulty we faced in trying to solve the analytical model, was in the calculation of $h_{m,n}(x)$ when $n\geq2$. 
As the reader can verify the final solution involves the use of special functions and it is far from trivial. 
The answer is:
\begin{align}
h_{m,n}(x) =  \frac{\alpha (\alpha\sqrt{2} \sigma_{n-m})^{m-1}}{(m-1)!}  I_{m,n}.
\end{align}  
The expression for $I_{m,n}$ turns out to be:
\begin{align}
I_{m,n} = 
\begin{cases}
  e^{\omega^2 - \psi^2 + (m-1) \ell n |\omega|}, \ \omega>0 \ \text{and} \  \omega^2 \gg 0\\
\frac{ e^{\omega^2 -\psi^2 } }{ 2\sqrt{\pi} }\left( \Gamma\left( \frac{m}{2} \right) M\left(\frac{1-m}{2}, \frac{1}{2}, -\omega^2 \right)  
+ 2|\omega| \Gamma\left( \frac{m+1}{2} \right) M\left(1-\frac{m}{2}, \frac{3}{2}, -\omega^2 \right)       \right), \ \omega > 0  \\
\frac{ e^{ -\psi^2 } }{2\pi}   \Gamma\left( \frac{m}{2} \right)   \Gamma\left( \frac{m+1}{2} \right) U\left( \frac{m}{2}, \frac{1}{2}, \omega^2  \right), \ \omega<0
\end{cases}
\nonumber
\end{align} 
where $M(a,b,z)$ and $U(a,b,z)$ are the \emph{confluent hypergeometric functions} of the first and second kind respectively~\cite{error}. 
The derivation can be found in the~\ref{app:Imn}. 
The variables $\omega$ and $\psi$ are given by the relations:
\begin{align}
\psi = & \frac{x-Q_{n-m}}{\sqrt{2}\sigma_{n-m}}, \\
\omega = & \frac{x-Q_{n-m} -\alpha\sigma^2_{n-m}}{\sqrt{2}\sigma_{n-m}}.
\end{align}  
Note that an asymptotic formula can be applied to $U(a,b,c)$ as well, for $\omega < 0$ and $\omega^2 \gg 0$, but this was not necessary in our work. 
Note also that for $n\geq10$ the $S_n(x)$ functions were approximated by symmetric gaussians to simplify the equations. 

\section{Generated data}
\label{sec:sim}

We remind the reader that the model, presented in the previous section, involves the following approximations:
\begin{enumerate}[i.]
\item All $G_n(x)$ functions for $n \geq 2$ were approximated with purely symmetric gaussians.
\item The $G_{n-m}(x)$ functions involved in the calculations of $h_{n,m}(x)$ were replaced with perfect gaussians and,  
\item $S_n(x)$ functions for $n\geq10$ were set equal to symmetric gaussians. 
\end{enumerate}
To understand the validity of these approximations, and assess the accuracy of the resulting model, we analyzed series of toy Monte Carlo data.  
PMT charge spectra were generated following the algorithm:
\begin{enumerate}[i.]
\item First, a number of PEs was thrown from a Poisson distribution of mean value $\mu$. 
\item For each PE, a charge was picked randomly from the SPE distribution of eq.~\eqref{eq:S}.
The total charge was calculated by summing the individual charges that each PE deposits.
\item A charge was thrown from the gaussian distribution of the pedestal and summed to the total charge obtained in step two.
\item The final, total charge was filled in a histogram.
\item This procedure was repeated $2.5\ 10^6$ times and a histogram of $2.5\ 10^6$ entries was produced.
\end{enumerate}

\begin{figure}[!t]
\centering
\includegraphics[width=12.0cm, height=8.20cm]{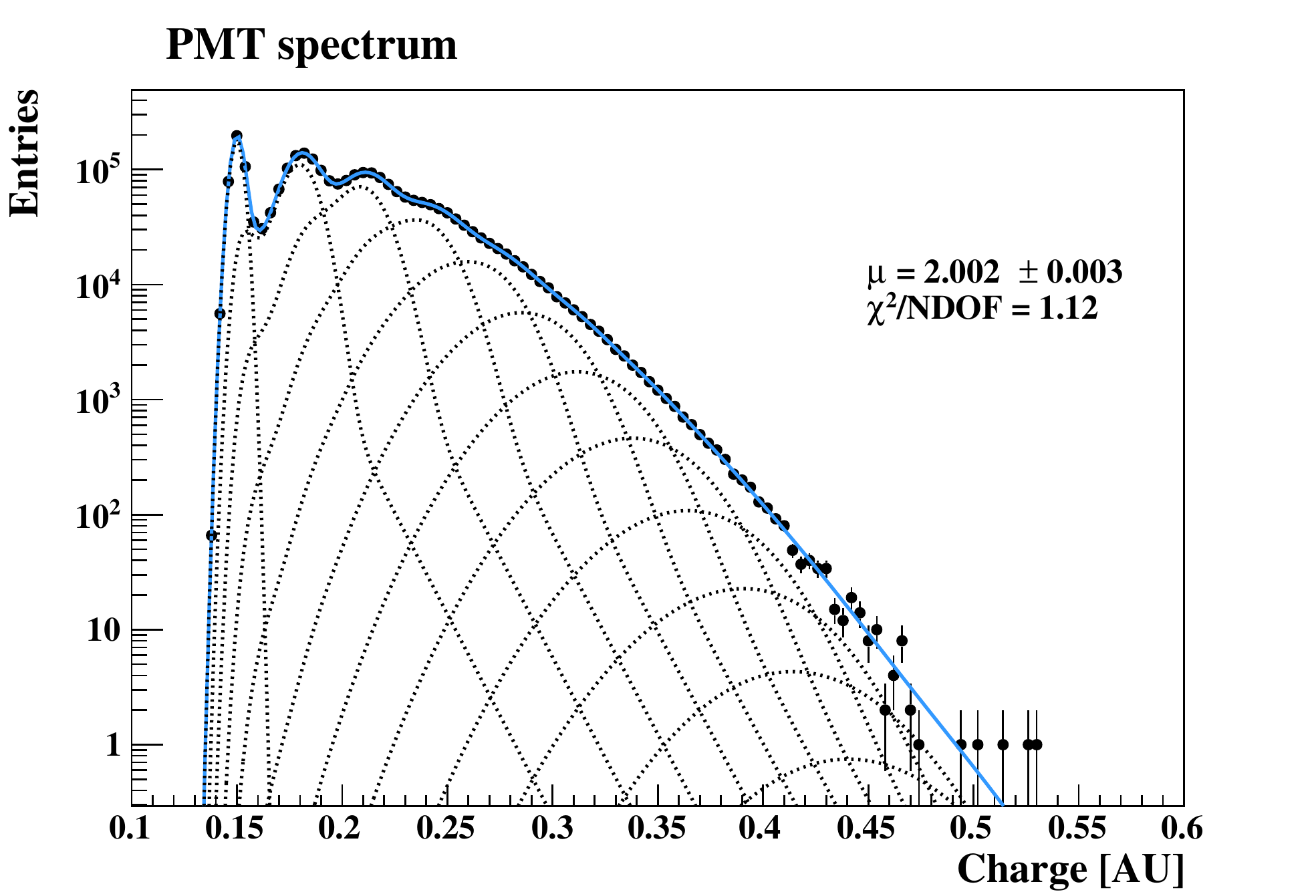} \\
\includegraphics[width=12.0cm, height=8.20cm]{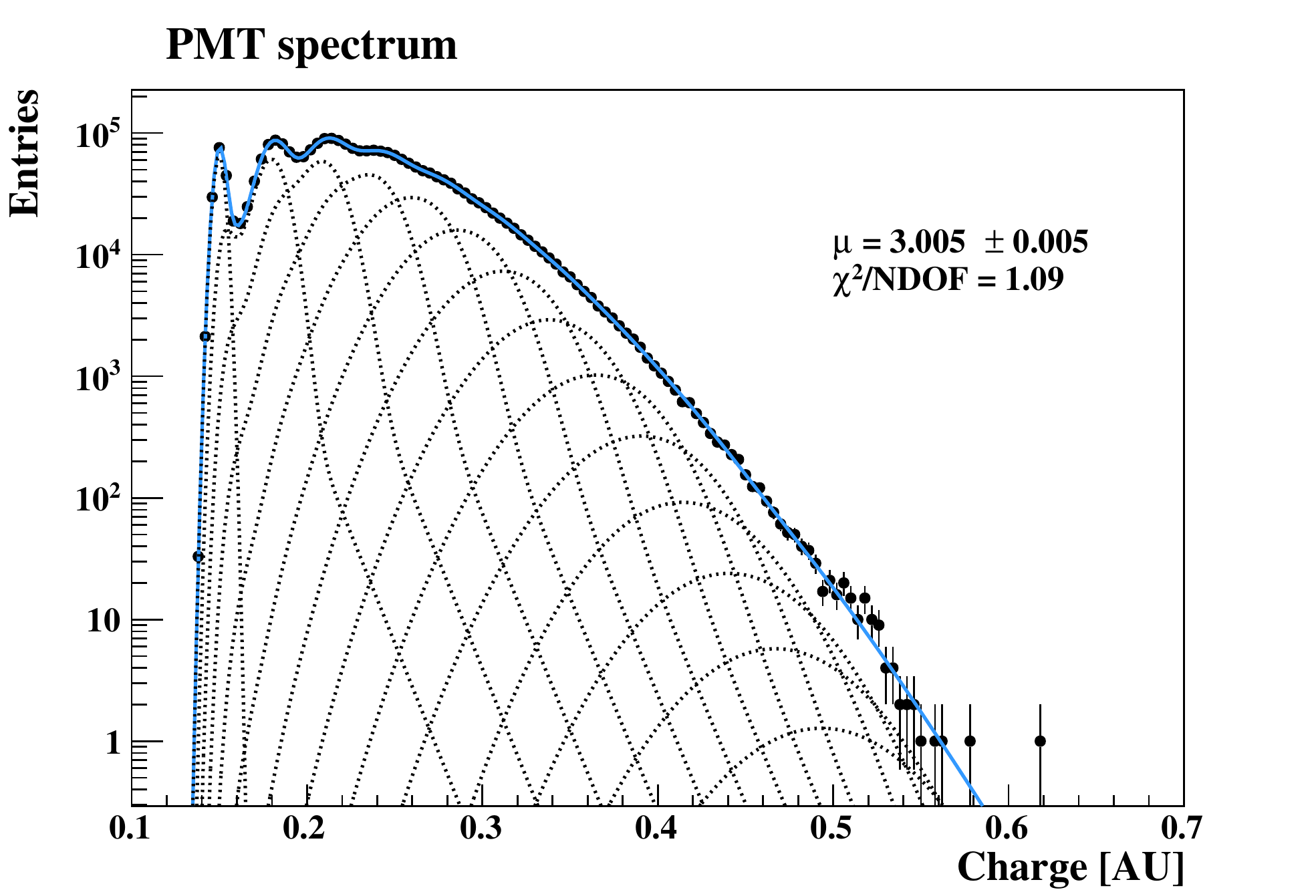} \\
\caption{ A few fits obtained with the analytical model. 
Black dots show the generated data and the azure line shows the best fit. The dashed lines show the contributions from the several PE peaks. 
The spectra generated with true values of $\mu$ equal to two (top) and three (bottom) are shown. }
\label{fig:spes}
\end{figure}
Examples of such distributions are shown in figure~\ref{fig:spes}. 
For all the histograms treated in this study, a simple gaussian fit was first performed around the maximum of the distribution to extract the mean ($Q_0$) and standard deviation ($\sigma_0$) of the pedestal. 
$Q_0$ and $\sigma_0$ were then fixed within a $\pm$ 2.5~\% tolerance. 
Additionally, the fit around the pedestal provided the normalization of the zero PE peak ($N_0$). 
An estimate of the mean number of PEs was given by the formula: 
\begin{align}
\mu \simeq - \ln \left(  \frac{N_0}{\ N_{tot}} \right),  
\label{eq:muestim}
\end{align}
and was input as an initial value to aid the minimizer. 
$N_{tot}$ is the total number of entries and eq.~\eqref{eq:muestim} is obtained from $P(0; \mu ) = e^{-\mu}$. 

The initial value of $Q$ was set to: 
\begin{align}
\frac{ \bar x - Q_0 }{\mu}
\label{eq:laestim}
\end{align}
where $\bar x$ is the mean of the charge histogram. 
Finally, $w$ was first set to 0.2 and limited between 0.0 and 0.6 (since for every well-functioning PMT the probability of badly amplified PEs cannot be larger than 60~\%).
Figure~\ref{fig:spes} shows a few examples of distributions fitted with the analytical model. 
The black dots show the generated data and the azure line shows the best fit. 
One can see that in both cases the fitted line follows closely the data and the $\chi^2$ over the number of degrees of freedom (NDOF) was always close to one. 
We also note that the software used for these studies exists in a public repository~\cite{git}.\footnote{%
A somewhat simpler and more straightforward example based solely on ROOT~\cite{root} has been committed in https://github.com/lkalousis/SimpleExample}

\begin{figure}[!t]
\centering
\includegraphics[width=12.0cm, height=8.25cm]{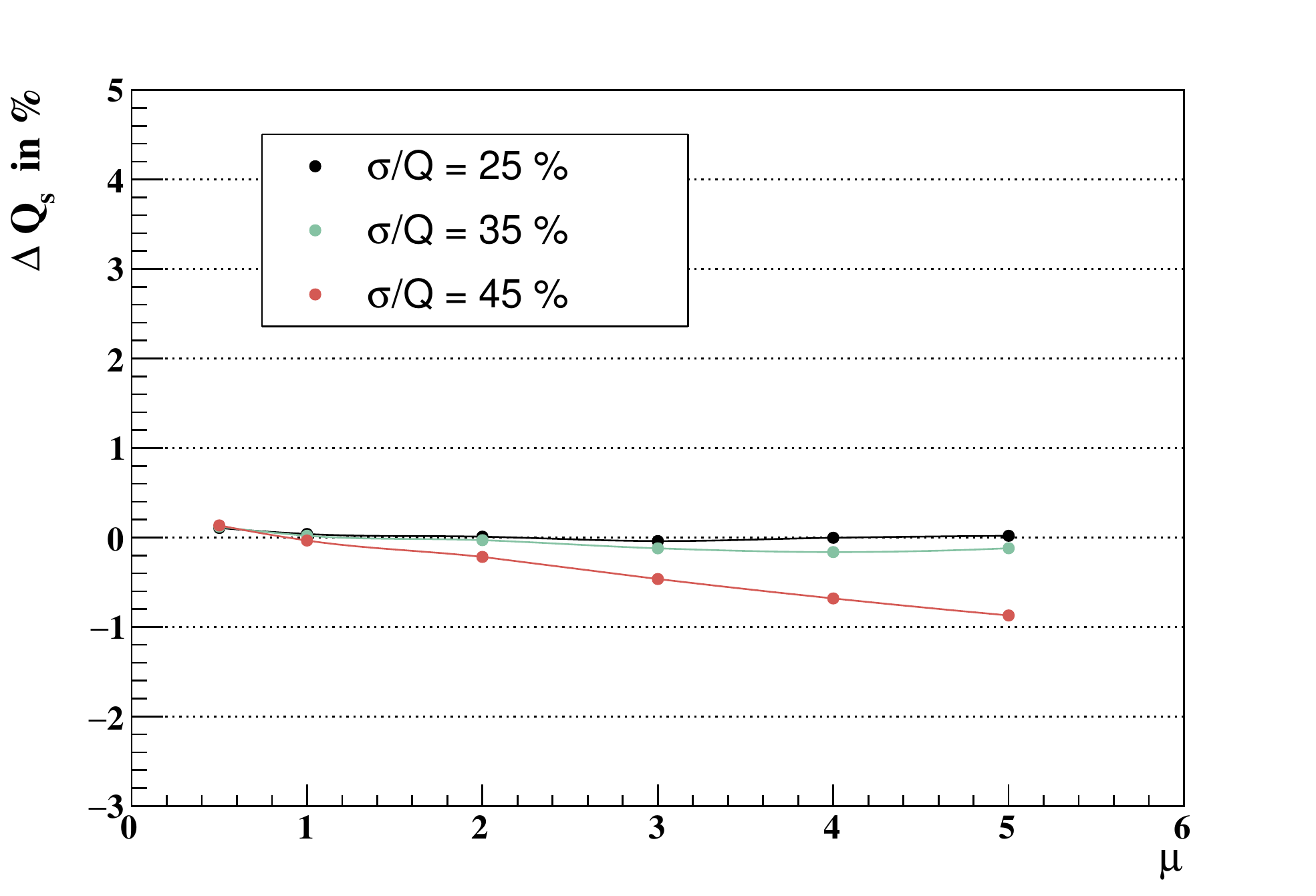} \\
\caption{ Accuracy of the analytical model for different values of $\mu$ and $\sigma/Q$. }
\label{fig:dev}
\end{figure}

Additionally, sets of one hundred toy data were generated for different values of $\mu$ inside the 0.5 -- 5.0 range. 
For each set of one hundred toys, the distribution of the relative deviation from the true gain ($Q_s$):
\begin{align}
\Delta Q_s = \frac{ Q_s^\prime - Q_s }{ Q_s }
\label{eq:laestim}
\end{align}
was plotted. We note that:
\begin{align}
 Q_s = \frac{w}{\alpha} + (1-w)Q_g,
\label{eq:laestim}
\end{align}
where $Q_g$ is the mean value of $g(x)$\footnote{The formulae of $Q_s$ and $Q_g$ together with their derivation can be found in ref.~\cite{me2}. } and $Q_s^\prime$ is the gain parameter returned by the minimizer.
Figure~\ref{fig:dev} shows the results of this exercise. 
The $x$ axis shows $\mu$ and the $y$ axis shows the mean value of the $\Delta Q_s$ distribution. 
Furthermore, to understand the effect of the truncation of the gaussian part of SPE,  data sets with different $\sigma/Q$ were analyzed. 
The results with $\sigma/Q$= 25 \% (black), 35 \% (green) and 45 \% (red) are shown in figure~\ref{fig:dev}. 
One can see that inside the $\sigma/Q$ = 25 -- 35 \% window the model extracts the gain with excellent accuracy. 
At the rather high value\footnote{And unrealistic for modern PMTs.} of $\sigma/Q$ = 45~\% there is a deviation from the true gain of $\sim$ 1.0 \% for $\mu = 5$.

\section{Calibration of the Hamamatsu R7081 photomultiplier}
\label{sec:res}

\subsection{Results}

\begin{figure}[!t]
\centering
\includegraphics[width=13.2cm, height=10.0cm]{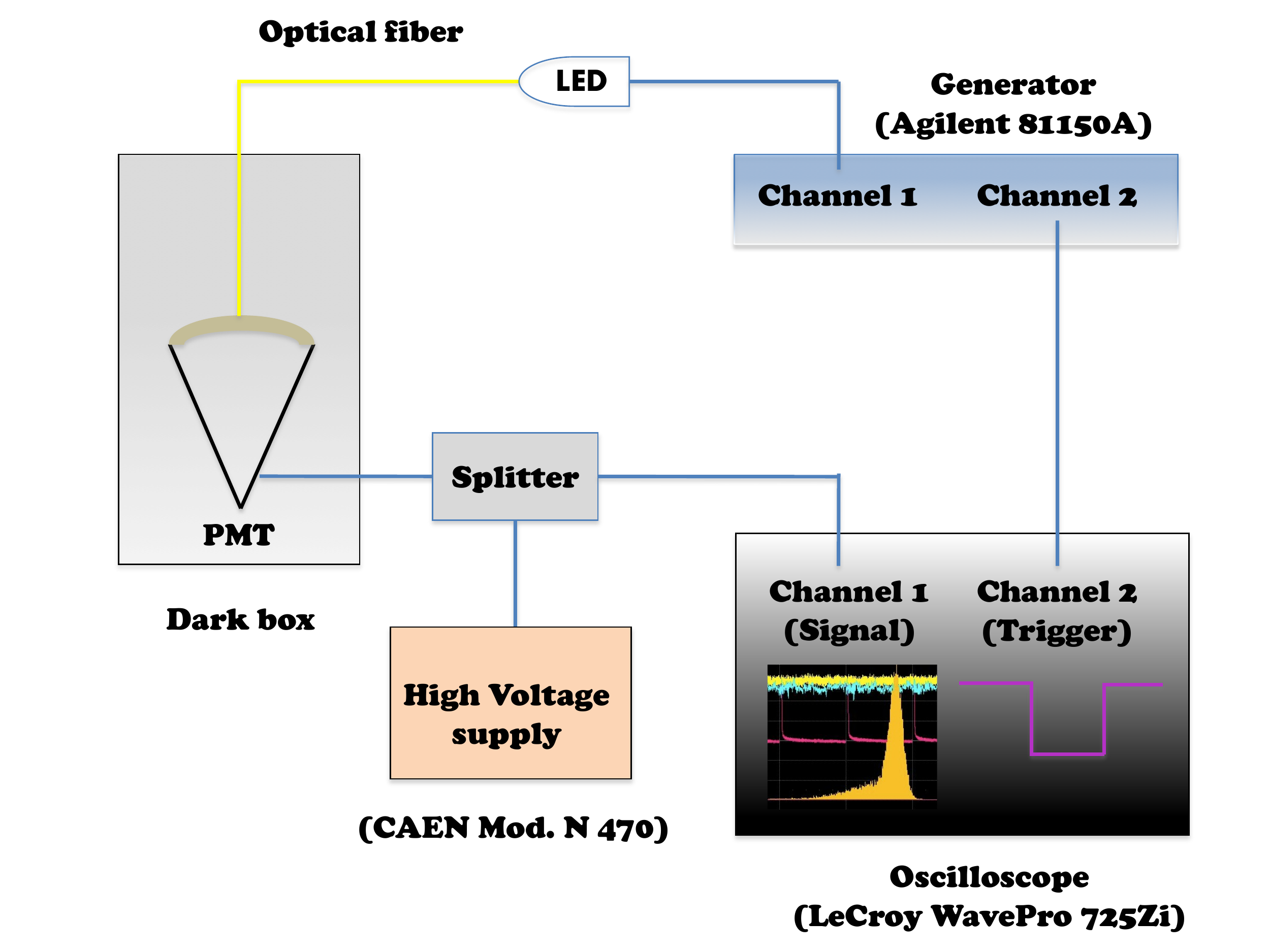} 
\caption{Pictorial representation of our PMT testing apparatus.}
\label{fig:setup}
\end{figure}

Several data sets were taken with a Hamamatsu R7081 PMT. 
The details of the experimental setup have been described before in ref.~\cite{me} and will not be repeated here. 
In a nutshell, the R7081 PMT was placed inside a dark, light-tight box. Light pulses were created by a light emitting diode (LED) connected to a fast pulse generator. 
The photons were carried inside the box through an optical fiber and the signals from the PMT were readout by a LeCroy WavePro 725Zi oscilloscope.   
The setup triggered on the generator's second, duplicated channel and all signals were recorded. Including those with no photons at all (pedestal). 
Figure~\ref{fig:setup} shows a simple schematic of the PMT testing apparatus. 

\begin{figure}[!t]
\centering
\includegraphics[width=11.80cm, height=8.25cm]{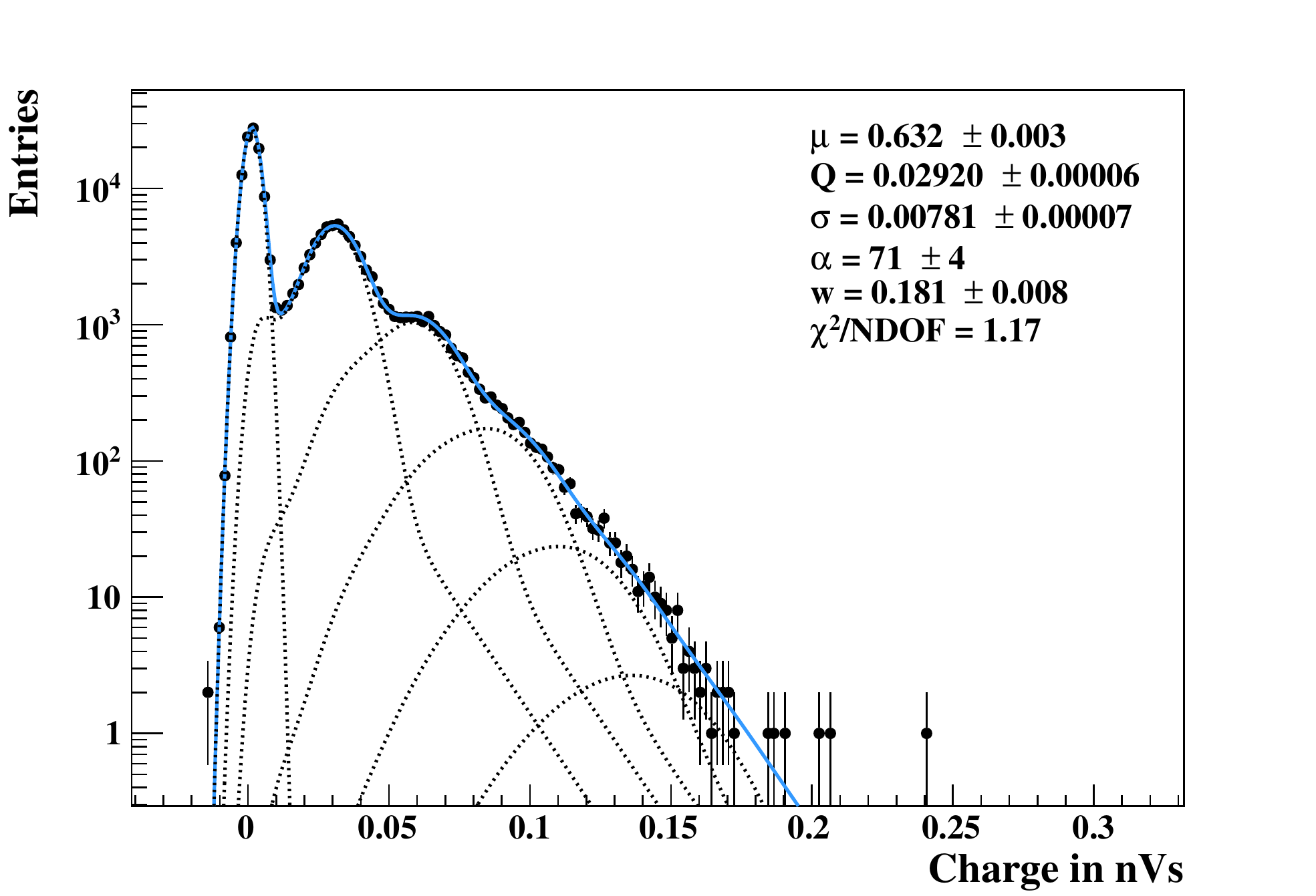} \\[1.5ex]
\includegraphics[width=11.80cm, height=8.25cm]{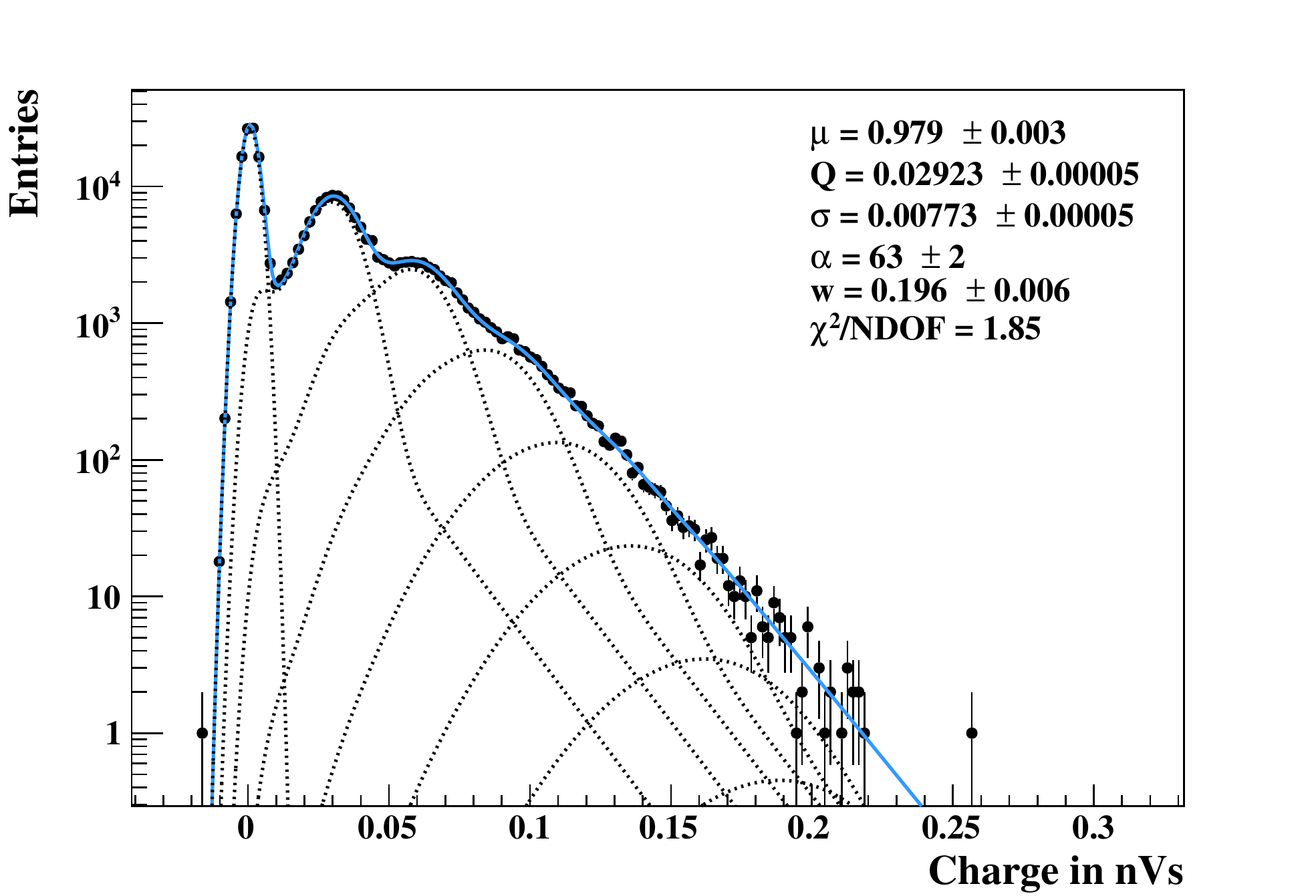} 
\caption{A few fits obtained using the analytical model of $S_R(x)$.  The data are shown in the black dots and the best fit curve is shown in azure line. }
\label{fig:spe}
\end{figure}
Figure~\ref{fig:spe} shows  a few fits obtained with the $S_R(x)$ model put forward in this article. 
Note that the unit of the $x$ axis is nanovolt times second (nVs), which is the unit of charge given by the LeCroy oscilloscope.
In particular, $Q$ and $\sigma$ are measured in nVs and $\alpha$ in nVs$^{-1}$. 
The best fit curve (azure line) follows closely the data and the $\chi^2$/NDOF is always close to one. 
To demonstrate that the fitting model is self-consistent and that extracts the correct $w$, $\alpha$, $Q$ and $\sigma$, several data were taken with varying LED intensity. 
The results of the analysis are shown in table~\ref{tab:money}. 
As the same PMT is used, the fit should output compatible values for all the parameters of $S(x)$ regardless of $\mu$. 
The pedestal might drift or change, depending on the stability of the setup, but the SPE response should stay the same. 

\begin{table}[t!]
\footnotesize
\centering
\begin{tabular}{| c  || c | c | c | c |}
\hline 
$\mu$  &  $w$ &  $\alpha$ (nVs$^{-1}$) &  $Q$ (nVs) &  $\sigma$ (nVs) \\[0.6ex] \hline\hline 
  0.544 $\pm$ 0.003 & 0.170 $\pm$  0.010 &  86 $\pm$ 10 & 0.02917  $\pm$ 0.00006 &  0.00789 $\pm$ 0.00008 \\ 
  0.632 $\pm$ 0.003 & 0.181 $\pm$  0.008 &  71 $\pm$ 4   & 0.02920  $\pm$ 0.00006 &  0.00781 $\pm$ 0.00007 \\ 
  0.769 $\pm$ 0.003 & 0.179 $\pm$  0.006 &  70 $\pm$ 3   & 0.02915  $\pm$ 0.00004 &  0.00784 $\pm$ 0.00005 \\ 
 0.979  $\pm$ 0.003 & 0.196 $\pm$  0.006 &  63 $\pm$ 2   & 0.02923  $\pm$ 0.00005 &  0.00773 $\pm$ 0.00005 \\ 
 1.357  $\pm$ 0.005 & 0.198 $\pm$  0.007 &  61 $\pm$ 2   & 0.02939  $\pm$ 0.00005 &  0.00777 $\pm$ 0.00006 \\ 
 2.014 $\pm$  0.008 & 0.197 $\pm$  0.007 &  61 $\pm$ 2   & 0.02935  $\pm$ 0.00006 &  0.00776 $\pm$ 0.00007 
\\[0.6ex] \hline\hline
\end{tabular}
\caption{Summary of the R7081 PMT calibration results.}
\label{tab:money}
\end{table}

Indeed, in table~\ref{tab:money} one can see that as $\mu$ becomes larger $w$, $\alpha$, $Q$ and $\sigma$ remain quite stable. 
A few further comments should be made. 
First, $Q$ and $\sigma$ can be obtained with much better accuracy than $w$ and $\alpha$. 
This is the case because the contribution of misamplified PEs is mainly constrained from the valley between the pedestal and the first PE peak. 
It appears that the minimizer has more difficulty to identify $w$ and $\alpha$ in this narrow region. 
Nonetheless, consistent results can be obtained. 

\begin{figure}[!t]
\centering
\includegraphics[width=12.25cm, height=8.5cm]{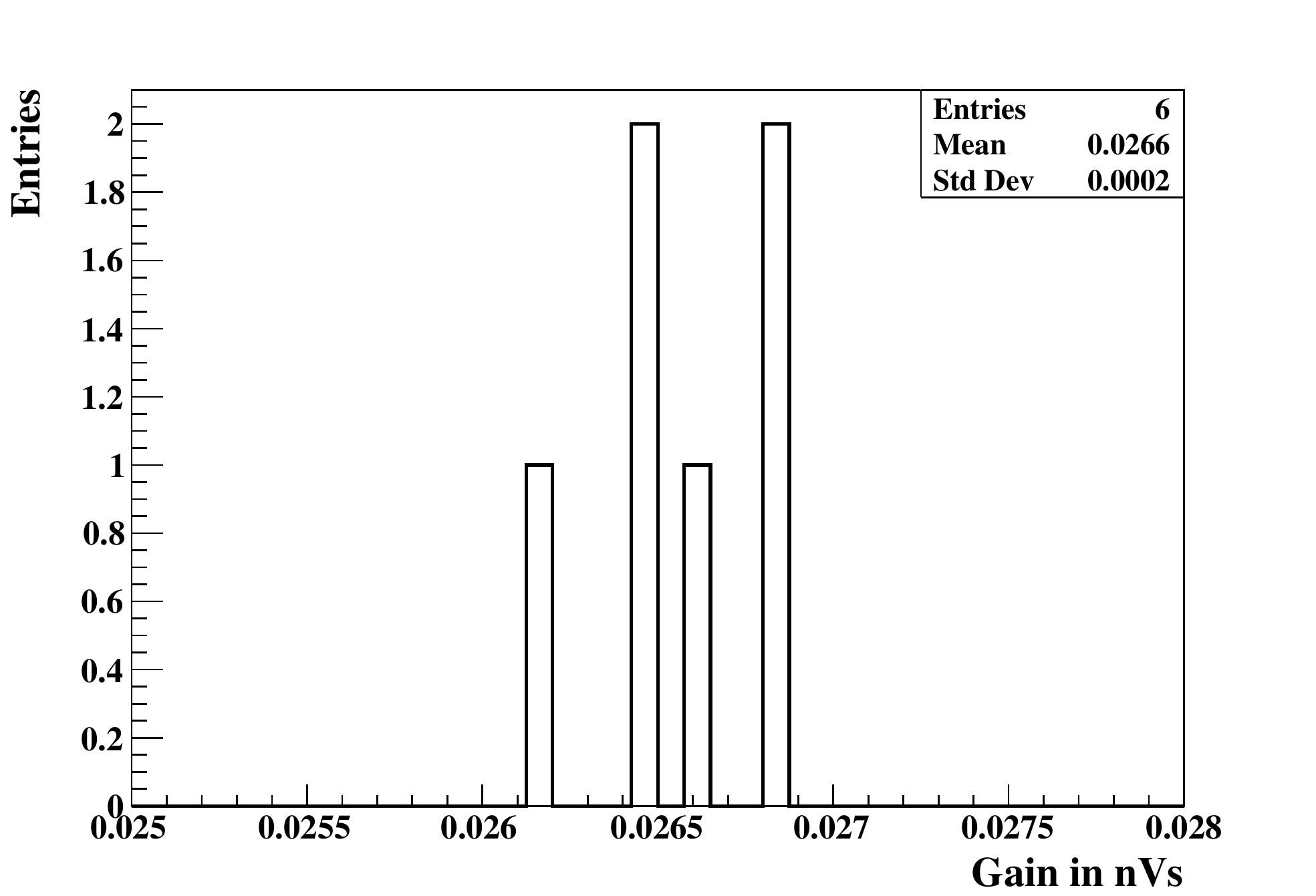} 
\caption{Distribution of the gain ($Q_s$) for the measurements compiled in table~\ref{tab:money}. }
\label{fig:g}
\end{figure}

Second, it is clear that correlations exist between the various parameters of $S(x)$. 
One can see that as $\mu$ increases there is a clear drift in the $w$ and $\alpha$ parameters. 
In particular, $w$ increases slightly and at the same time $\alpha$ decreases. 
Nonetheless, this drift is small in magnitude (almost within errors) and a straightforward calculation of the gain ($Q_s$), using eq.~\eqref{eq:laestim}, gives results that are consistent. 
Figure~\ref{fig:g} shows the one-dimensional distribution of $Q_s$ for those measurements compiled in table~\ref{tab:money}. 
The relative standard deviation of the distribution in figure~\ref{fig:g} is very small, highlighting that one can determine $Q_s$ with a $\sim$1\% accuracy. 
These are important results, indicating that the analytical model can be useful in the absolute calibration of PMTs providing precise results in a vast range of PEs.

\subsection{Comparison with the Discrete Fourier Transform approach}

The data set of table~\ref{tab:money} has been analyzed before in ref.~\cite{me2}. 
For that purpose, a numerical method based on the Discrete Fourier Transform (DFT) was employed~\cite{me}. 
In this section we present a comparison of the results obtained through analytical model and the DFT methods. 
 Table~\ref{tab:comp} shows the gain ($Q_s$) obtained through the DFT method and the analytical model. 
The first row shows the Poisson mean $\mu$ extracted by the fitter for reference purposes. 
One can observe that as $\mu$ increases the gain remains quite stable and the results between the DFT and the analytical model are almost identical. 
We should also note that a drift in the gain is observed (owing to the correlations highlighted in previous section) but this is too small in magnitude.
It is a deviation of $\sim$ 2$\sigma$ and it is well within the measurement errors.    

\begin{table}[t!]
\centering
\begin{tabular}{| c  || c | c |}
\hline
$\mu$                      & $Q_s$, DFT method (nVs)  	       & $Q_s$, analytical model (nVs)           \\[0.6ex] \hline\hline
0.544 $\pm$ 0.003  & 0.0262 	$\pm$ 0.0003			       	       & 0.0262 	$\pm$ 0.0003					     				\\
0.632 $\pm$ 0.003  & 0.0265	$\pm$ 0.0003				       & 0.0265	$\pm$ 0.0003				  	    				\\
0.769 $\pm$ 0.003  & 0.0265	$\pm$ 0.0003				       & 0.0265	$\pm$ 0.0003					  	    				\\
0.979 $\pm$ 0.003  & 0.0266	$\pm$ 0.0002				       & 0.0266	$\pm$ 0.0002				    	    				\\
1.357 $\pm$ 0.005  & 0.0268	$\pm$ 0.0003				       & 0.0268    $\pm$ 0.0003						  	    				\\
2.014 $\pm$ 0.008  & 0.0268	$\pm$ 0.0003				       & 0.0268	$\pm$ 0.0003				  			
\\[0.6ex] \hline\hline
\end{tabular}
\caption{Comparison between the analytical model and DFT methods.}
\label{tab:comp}
\end{table}

Furthermore, in ref.~\cite{me2} we have analyzed the same data set using the numerical integration method mentioned in section~\ref{sec:intro}.  
To simplify the calculations only the first PE peak was worked out numerically and higher peaks were approximated with perfectly symmetric gaussians. 
It was shown that inside the $\mu$ = 0.5 - 2.0 window the numerical integration gave the same results as the DFT approach. 
So, we can expect all three techniques to provide the same figures of tables 1 and 2. 
Additionally, it should be mentioned that the numerical integration gave a slightly worse $\chi^2$/NDOF and took more time to analyze this data set. 
Calculating more peaks numerical will improve $\chi^2$/NDOF but at the same time will increase the execution time rendering the method almost inapplicable. 
More details can be found in ref.~\cite{me2}.

\section{Outlook}
\label{sec:outro}
In this article we have presented an analytical model that can be used for the calibration of PMTs whenever the SPE response can be parameterized by eq.~\eqref{eq:S}.  
Note that the SPE charge amplification of the full dynode chain for the R7081 PMT has a very symmetrical shape and it can be modeled by a truncated gaussian.  
In several cases, as for instance in the R1408 Hamamatsu PMT, the SPE response has a rather asymmetrical shape and the formulae given in this publication seize  to apply.  
It has been argued that this is due to venetian-blind type of the dynode array.   
In this example a gamma distribution is to be preferred~\cite{me}. 

The various mathematical formulae were derived with great detail and attention was paid to showcase the approximations needed for the final solution. 
We should emphasize that while the equations might seem, at first glance, both complicated and unnecessarily elaborate, 
we find that the model is quite simple and that it can be employed with the same efficacy as the already establish approaches. 
Data from a R7081 PMT were analyzed and the results were in excellent agreement with those obtained using the DFT method. 
In particular, the gain parameter of the PMT was remarkably stable within the $\mu$~$\sim$~0.5 - 2.0 plateau. 
The code used for these investigations exists in a public repository~\cite{git}.  

We would like to emphasize that the main thrust of our efforts was directed towards two specific objectives:
\begin{enumerate}[i.]
\item First, we wanted to derive all the necessary mathematical formulae in a clear and pedagogical manner that everybody understands and,
\item Second, we wanted to demonstrate that the model can extract the gain parameter in a rigorous way, similar to those of already established methods. 
In this respect, we showed that it returns figures consistent with the numerical DFT technique.  
\end{enumerate}
Of course, the procedure outlined in this publication can be used to study further properties of PMTs with gaussian single photoelectron response function. 
For instance, it could be employed to probe the photoelectron detection efficiency (PDE) of PMTs with characteristics similar to the Hamamatsu R7081. 

Finally, we should also mention that the machinery developed in this article comprises the first analytical solution of the Dossi model. 
As an example, we can refer to the Darkside paper, ref.~\cite{dark}, where the equations were solved numerically for the first two PE peaks 
and higher peaks were approximated with symmetric gaussians. 
This approach is accurate only for small values of $\mu$, where the non-gaussian tails of $S^{(n)}(x)$ can be neglected. 
In striking contrast, the model of this article can incorporate higher PE corrections, in the software, with only a few lines of code. 
Additionally, a solution exists in a closed form and a single spectrum can be analyzed in just a few seconds of execution time. 
We believe that this is an important improvement over the existing approximate methods.


\appendix


\section{$\int_{ x_1 }^{x_2}  e^{-a x^2 +b x +c } \ dx$ integral}
\label{app:int}

The integral, 
\begin{align}
I = \int_{ x_1 }^{x_2}  e^{-a x^2 +b x +c } \ dx,
\end{align}
occurs frequently in calculations involving $S_R(x)$ functions with SPE responses parameterized with gaussians, and so we think that it is rather instructive to derive its formula in this appendix. 
\begin{align}
I = & \int_{ x_1 }^{x_2}  e^{-a x^2 +b x +c } \ dx \nonumber \\
  = & e^c \int_{ x_1 }^{x_2}  e^{-a ( x^2 - \frac{b}{a} x )} \ dx  \nonumber \\
  = & e^{\frac{b^2}{4a}+c} \int_{ x_1 }^{x_2}  e^{-a ( x - \frac{b}{2a} )^2} \ dx. 
\end{align}
Using the substitution: 
\begin{align}
u = \sqrt{a} ( x - \frac{b}{2a} )  
\end{align}
$I$ becomes,
\begin{align}
I = &  \frac{1}{\sqrt{a}} e^{\frac{b^2}{4a}+c} \int_{ \frac{2a x_1 -b }{2\sqrt{a}} }^{ \frac{2a x_2 -b }{2\sqrt{a}} }  e^{-u^2 } \ du \nonumber \\
  = & \frac{1}{\sqrt{a}} e^{\frac{b^2}{4a}+c} \left(   \int_{ 0 }^{ \frac{2a x_2 -b }{2\sqrt{a}} }  e^{-u^2 } \ du -  \int^{ \frac{2a x_1 -b }{2\sqrt{a}} }_{ 0 }  e^{-u^2 } \ du\right)  \nonumber \\
  = &  \frac{1}{2}  \sqrt{\frac{\pi}{a} } e^{\frac{b^2}{4a}+c} \left[ \text{erf}\left({ \frac{2a x -b }{2\sqrt{a}} }\right) \right]^{ x_2 }_{ x_1 }.
\end{align}
In the last line we made use of the definition:
\begin{align}
 \text{erf}(x) = \frac{2}{\sqrt{\pi}} \int_0^x e^{-t^2} dt. 
 \end{align}
 Now, if $x_2\rightarrow +\infty$ the integral simplifies to:
 \begin{align}
I = &  \frac{1}{2}  \sqrt{\frac{\pi}{a} } e^{\frac{b^2}{4a}+c} \left( 1 - \text{erf}\left({ \frac{2a x_1 -b }{2\sqrt{a}} }\right) \right) \nonumber \\
  = & \frac{1}{2}  \sqrt{\frac{\pi}{a} } \ e^{\frac{b^2}{4a}+c} \ \text{erfc}\left({ \frac{2a x_1 -b }{2\sqrt{a}} }\right). \label{eq:intg}
\end{align}

\section{Formulae for $(f*B)(x)$ and $(g*B)(x)$}
\label{app:sr1}

\subsection*{Case of $(f*B)(x)$ }

We start with a direct calculation of the $(f*B)(x)$ convolution integral:
\begin{align}
(f*B)(x) = & \frac{\alpha}{\sqrt{2\pi}\sigma_0}   \int^{+\infty}_{-\infty} e^{ -\frac{(x-Q_0-t)^2}{2\sigma_0^2} -\alpha t } H(t) dt \nonumber \\
             = & \frac{\alpha}{\sqrt{2\pi}\sigma_0}  \int^{+\infty}_{0} e^{ -\frac{(x-Q_0-t)^2}{2\sigma_0^2} -\alpha t } dt.
\end{align}
Through the substitution:
\begin{align}
u = \frac{t+Q_0-x}{\sqrt{2}\sigma_0}  
\end{align}
the expression becomes:
\begin{align}
(f*B)(x) = & \frac{\alpha }{\sqrt{\pi}}   \int^{+\infty}_{ \frac{Q_0-x}{\sqrt{2}\sigma_0}   } e^{ -u^2 -\alpha( \sqrt{2}\sigma_0 u + x - Q_0 ) } du \nonumber \\
            = & \frac{\alpha }{\sqrt{\pi}} e^{-\alpha(x-Q_0)} \int^{+\infty}_{ \frac{Q_0-x}{\sqrt{2}\sigma_0}   } e^{ -u^2 - \sqrt{2}\alpha\sigma_0 u  } du  \nonumber \\
            = & \frac{\alpha }{2} e^{\frac{\alpha^2\sigma_0^2}{2}} e^{-\alpha(x-Q_0)} \ \text{erfc}\left(    \frac{Q_0 + \alpha\sigma_0^2 -x }{\sqrt{2}\sigma_0} \right).
\end{align}
In the last line we employed the integral identity~\eqref{eq:intg}.

\subsection*{Case of $(g*B)(x)$ }

In a similar way the calculation of  $(g*B)(x)$ proceeds. 
\begin{align}
(g*B)(x) = & \frac{1 }{g_N}  \frac{1 }{\sqrt{2\pi}\sigma} \frac{1 }{\sqrt{2\pi}\sigma_0} \int^{+\infty}_{ -\infty  }  e^{-\frac{(t-Q)^2}{2\sigma^2} - \frac{(x-t-Q_0)^2}{2\sigma_0^2}} H(t) dt
\end{align}
\begin{align}
I = & \int^{+\infty}_{ -\infty  }  e^{-\frac{(t-Q)^2}{2\sigma^2} - \frac{(x-t-Q_0)^2}{2\sigma_0^2}} H(t) dt \nonumber \\
  = & \int^{+\infty}_{ 0  }  e^{-\frac{(t-Q)^2}{2\sigma^2} - \frac{(x-t-Q_0)^2}{2\sigma_0^2}} dt
\end{align}
With the substitutions,
\begin{align}
u = \frac{t-Q}{\sigma} \ \  \text{and}\  \ Q^\prime = x - Q - Q_0,
\end{align}
$I$ takes the form:
\begin{align}
I = \ & \sigma \int^{+\infty}_{ -\frac{Q}{\sigma}  }  e^{-\frac{u^2}{2} - \frac{(Q^\prime-\sigma u )^2}{2\sigma_0^2}} du \nonumber \\
  = \ & \sigma \int^{+\infty}_{ -\frac{Q}{\sigma}  } e^{-\frac{\sigma^2+\sigma_0^2}{2\sigma_0^2 }u^2 + \frac{Q^\prime\sigma}{\sigma_0^2}u - \frac{Q^{\prime 2}}{2\sigma_0^2}  } du \nonumber \\
  = \ &  \frac{1}{2} \sqrt{2\pi} \sigma \sigma_0 \frac{1}{\sqrt{\sigma^2 +\sigma_0^2}} \ e^{ -\frac{(x-Q_0-Q)^2}{2( \sigma_0^2 + \sigma^2 )}} 
  \text{erfc} 
  \left(    \frac{ Q_0\sigma^2 -Q\sigma_0^2 -x \sigma^2  }{\sqrt{2} \sigma_0\sigma\sqrt{\sigma_0^2 + \sigma^2} }         \right). 
\end{align}
Again we made use of the integral of~\ref{app:int}. 
Plugging $I$ into the initial equation for $(g*B)(x)$ we are left with the solution: 
\begin{align}
(g*B)(x) = \frac{1}{2} \frac{1}{g_N}  \frac{1}{\sqrt{2\pi}} \frac{1}{\sqrt{\sigma_0^2 +\sigma^2}}  \ e^{ -\frac{(x-Q_0-Q)^2}{2( \sigma_0^2 + \sigma^2 )}} 
\text{erfc} 
  \left(    \frac{ Q_0\sigma^2 -Q\sigma_0^2 -x \sigma^2  }{\sqrt{2} \sigma_0\sigma\sqrt{\sigma_0^2 + \sigma^2} }         \right). 
\end{align}

\section{Calculation of $h_{m,n}(x)$}
\label{app:Imn}

We first write down the formula for $f_m(x)$:
\begin{align}
f_m(x) =  \frac{\alpha (\alpha x )^{m-1}}{(m-1)!} e^{-\alpha x } H(x),
\end{align}
which one can easily prove by induction. 
$h_{m,n}(x)$ becomes:
\begin{align}
h_{m,n}(x) = & \frac{ \alpha^{m}}{(m-1)!} \frac{1}{\sqrt{2\pi}\sigma_{n-m}} \int_{-\infty}^{+\infty} t^{m-1} e^{ -\frac{ (x-t -Q_{n-m})^2}{2\sigma^2_{n-m}}-\alpha t} H(t) dt \nonumber \\
                  = & \frac{ \alpha^{m}}{(m-1)!} \frac{1}{\sqrt{2\pi}\sigma_{n-m}} \int_{0}^{+\infty} t^{m-1} e^{ -\frac{ (x-t -Q_{n-m})^2}{2\sigma^2_{n-m}}-\alpha t} dt.
\end{align}
Note that $G_{n-m}(x)$ was replaced with a symmetric gaussian. 
Completing the square in the exponential one has:
\begin{align}
h_{m,n}(x) = & \frac{ \alpha^{m}}{(m-1)!} \frac{1}{\sqrt{2\pi}\sigma_{n-m}} \int_{0}^{+\infty} t^{m-1} e^{  \omega^2  -\psi^2  -\left( \frac{t}{\sqrt{2}\sigma_{n-m}} - \omega \right)^2} dt,
\end{align}
with 
\begin{align}
\psi = & \frac{x-Q_{n-m}}{\sqrt{2}\sigma_{n-m}}, \\
\omega = & \frac{x-Q_{n-m} -\alpha\sigma^2_{n-m}}{\sqrt{2}\sigma_{n-m}}.
\end{align}  
The integral:
\begin{align}
I =  e^{  \omega^2  -\psi^2 }\int_{0}^{+\infty} t^{m-1} e^{  -\left( \frac{t}{\sqrt{2}\sigma_{n-m}} - \omega \right)^2} dt,
\end{align}
becomes with the substitution:
\begin{align}
u = \frac{t}{\sqrt{2}\sigma_{n-m}},
\end{align}
\begin{align}
I =  (\sqrt{2}\sigma_{n-m})^m e^{  \omega^2  -\psi^2 }\int_{0}^{+\infty} u^{m-1} e^{  -\left( u - \omega \right)^2} du.
\end{align}
Now, the integral:
\begin{align}
I^\prime = \int_{0}^{+\infty} u^{m-1} e^{  -\left( u - \omega \right)^2} du,
\end{align}
was solved using the \emph{Mathematica} online software~\cite{math} and the result is:
\begin{align}
I^\prime = \frac{1}{2} e^{-\omega^2} 
\left(  \Gamma\left( \frac{m}{2} \right) M\left( \frac{m}{2}, \frac{1}{2}, \omega^2 \right)  + 2\omega \Gamma\left( \frac{m+1}{2} \right) M\left(\frac{m+1}{2}, \frac{3}{2}, \omega^2  \right)  \right).  
\end{align}
Finally, and using the formulae for $I$ and $I^\prime$, $h_{m,n}(x)$ takes the form:
\begin{align}
h_{m,n}(x) =  \frac{ \alpha (\alpha\sqrt{2} \sigma_{n-m})^{m-1}}{(m-1)!}  I_{m,n}
\end{align} 
where $I_{m,n}$  is given by the equation:
\begin{align}
I_{m,n} = \frac{ e^{-\psi^2} }{ 2\sqrt{\pi} }\left( \Gamma\left( \frac{m}{2} \right) M\left( \frac{m}{2}, \frac{1}{2}, \omega^2 \right)  + 2\omega \Gamma\left( \frac{m+1}{2} \right) M\left(\frac{m+1}{2}, \frac{3}{2}, \omega^2 \right) \right).  
\end{align}

\subsection*{$I_{m,n}$ case I: $\omega> 0$ and $\omega^2 \gg 0$ }

We first treat the case where $\omega$ is positive and $\omega^2$ is far greater than zero, $\omega^2 \gg 0$.  
We take advantage of the asymptotic formula for $M(a,b,z)$ when $z \gg 0$:
\begin{align}
M(a,b,z) \simeq \frac{\Gamma (b)}{\Gamma(a)} \  e^z z^{a-b}.
\end{align}
We then have for $I_{m,n}$:
\begin{align}
 I_{m,n} \simeq e^{\omega^2 - \psi^2 + (m-1) \ell n |\omega|}   .
\end{align}

\subsection*{$I_{m,n}$ case II: $\omega >0$}

We use Kummer's transformation:
\begin{align}
M(a,b,z) = e^z M(b-a,b,-z)
\end{align}
and $I_{m,n}$ becomes:
\begin{align}
I_{m,n} = \frac{ e^{\omega^2 -\psi^2 } }{2\sqrt{\pi}} \left(  \Gamma\left( \frac{m}{2} \right) M\left(\frac{1-m}{2}, \frac{1}{2}, -\omega^2 \right)  \nonumber \right. \\ \left. 
+ 2|\omega| \Gamma\left( \frac{m+1}{2} \right) M\left(1-\frac{m}{2}, \frac{3}{2}, -\omega^2 \right)     \right).
\end{align}

\subsection*{$I_{m,n}$ case III: $ \omega <0$}

For negative values of $\omega$ one can exploit the relation:
\begin{align}
U(a,b,z) = \frac{\Gamma(1-b) }{\Gamma(a+1-b)}M(a,b,z) + \frac{\Gamma(b-1) }{\Gamma(a)} z^{1-b} M(a+1-b, 2-b, z )
\end{align}
and write $I_{m,n}$ as:
\begin{align}
I_{m,n} = \frac{ e^{ -\psi^2 } }{ 2\pi }\Gamma\left( \frac{m}{2} \right)   \Gamma\left( \frac{m+1}{2} \right) U\left( \frac{m}{2}, \frac{1}{2}, \omega^2  \right). 
\end{align}
We note that a similar  asymptotic relation for $U(a,b,z)$ can be written for $\omega<0$ and $\omega^2 \gg 0$. 




\section*{Acknowledgements}

The data used in this publication were taken in the laboratory of M. Dracos and we wish to thank him for allowing us to use them for the purposes of this communication. 
We also wish to thank C. Bobeth for encouraging us to continue and complete this study.

\end{document}